\documentclass{article}
\usepackage{graphicx} 
\usepackage[natbibapa]{apacite}

\title{Imposed mesoscale moisture heterogeneity enhances shortwave cloud radiative effects in large-eddy simulations of trade cumuli}
\author{M. van Driel$^{1,2}$, C. C. van Heerwaarden$^1$, M. Janssens$^{1,2}$}

\begin{document}
\maketitle

1. Wageningen University \& Research, Wageningen, The Netherlands

2. Delft University of Technology, Delft, The Netherlands

Corresponding author: M. van Driel, marloes.vandriel@wur.nl

\textbf{Key points}:

\begin{itemize}
\item We compare trade cumuli in 512 km idealized large eddy simulations with 144 and 250 km initial moisture perturbations to a homogeneous case
\item The initial perturbations grow in size and amplitude, leading to a permanently altered cloud, that reflects 4-9 W/m$^2$ more solar radiation
\item Therefore, small frequency changes of variance advection into the trades with warming could greatly alter the trade cumulus feedback
\end{itemize}

\begin{abstract}
The trade cumulus cloud feedback remains uncertain. We use $512 \times 512$ km$^2$ idealized Large Eddy Simulations (LESs) to investigate whether the import of mesoscale moisture variance to the trades can affect this feedback. Specifically, we compare a 'Control' LES and three 'Perturbed' LESs, where $\geq$144 km moisture variance is added while keeping the domain-mean moisture constant. The Perturbed cases develop significantly larger ($\sim$300 km) clouds than the Control case ($\sim$100 km), contain more stratiform clouds, and hence develop a larger daily-mean cloud cover and shortwave cloud radiative effect than the Control case (4.4, 6.6, and 8.7 W/m$^2$). Thus, the frequency of moisture variance advected into the trades with warming could constitute a potentially important, overlooked contribution to the cumulus feedback. We therefore argue that quantifying the shallow cloud feedback requires a quantitative description of the moisture variance–radiation relationship, and a better understanding of mesoscale moisture variance advection into the trades. 

\end{abstract}

\section*{Plain Language Summary}
Shallow trade-wind cumulus clouds reflect sunlight back to space, and they thus cool the earth. They occur in many different patterns. These clouds and their patterns respond to and feedback on climate change. However, the extent of this feedback is still uncertain. We create four patterns in idealized computer experiments resembling areas of $512 \times 512$ km$^2$: one usual (Control) and three 'Perturbed' experiments with enhanced moisture in the middle of the domain, created such that the domain-mean moisture is kept constant. The control experiment produces clouds of around 100 km, while the Perturbed experiments produces a single cloud of around 300 km. Moreover, the Perturbed experiments have a larger cloud cover and therefore more reflection of sunlight. If perturbations would be advected into the trades more often in future (due to climate change), this would lead to more cooling of the earth by these clouds (and thus a smaller feedback, or vice versa. Therefore, understanding these patterns and their relation to perturbations is crucial for quantifying the shallow cloud feedback.

\section{Introduction}
\label{sec:intro}
The shallow trade-wind cumulus cloud regime covers around 20\% of the planet \citep{Myers2021ObservationalSensitivity, Scott2020ObservedOceans}, and these clouds effectively cool the earth by reflecting incoming solar radiation without reflecting a lot of earth's warmth. Because of this large coverage, a small change in these clouds may have a large effect on the global incoming solar radiation. These clouds tend to organize in mesoscale (10 to 1000 km) patterns and we do not fully understand the organizing mechanisms, nor whether the patterns play a critical role in the climate feedback \citep{Nuijens2019BoundaryClimate, Vial2017MechanismsReview}.

\begin{figure}[h!]
    \centering
    \includegraphics[width=1\textwidth]{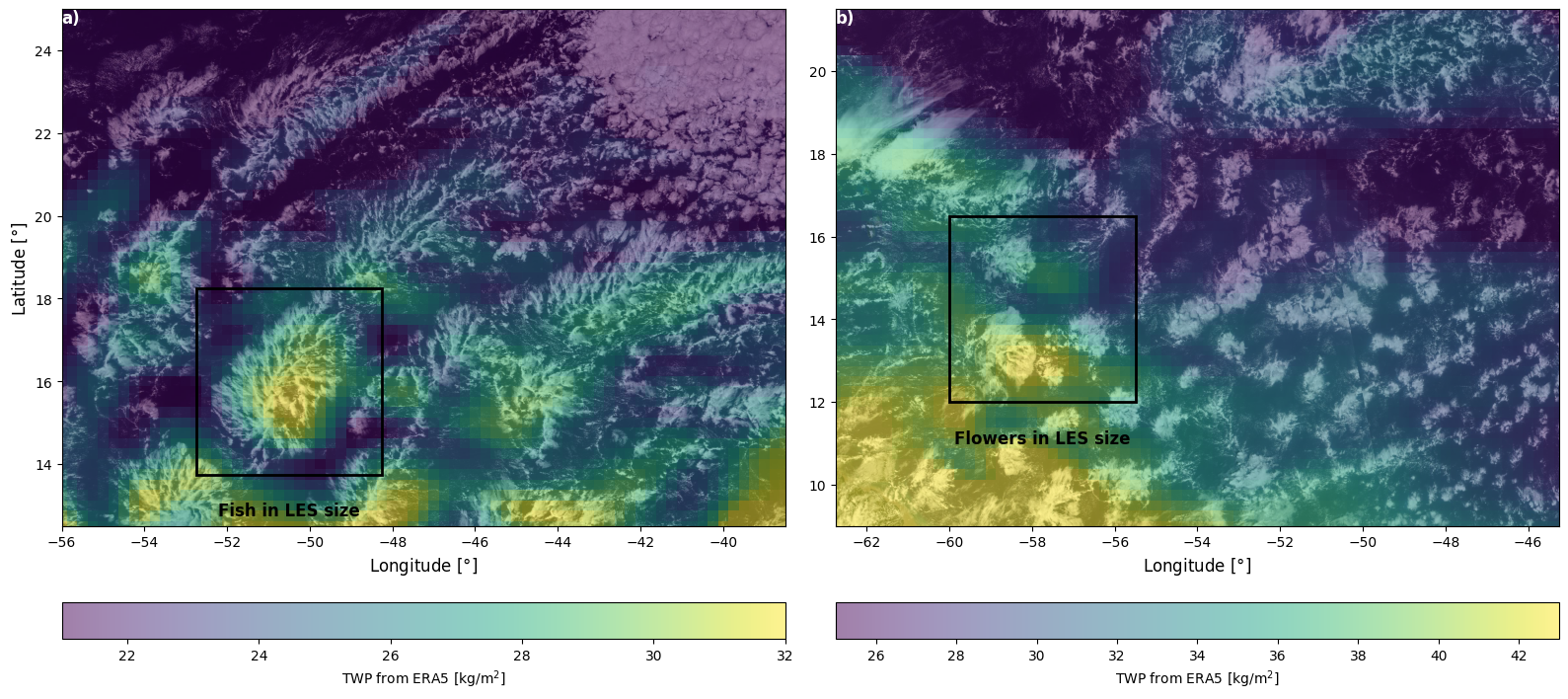}
    \caption{Two cloud scenes (white) containing a) several Fish clouds and b) several Flower clouds obtained from NOAA-21/VIIRS overlaid with the Total Water Path (TWP) from ERA5 (color scale).}
    \label{fig:Fish-sat}
\end{figure}

The efforts in recent years to understand these organizing mechanisms have primarily focused on spontaneous growth of cloud structures, through idealized Large Eddy Simulations (LESs). For example, such efforts focused on fine small scale cloud structures (sometimes called Sugar clouds) \citep{Janssens2023NonprecipitatingGrowth}, aggregated clusters (sometimes called Flower clouds, Figure \ref{fig:Fish-sat}b) \citep{Narenpitak2021FromATOMIC, Narenpitak2023TheDust, Kazil2024OnOrganization}, and cloud arcs (sometimes called Gravel clouds) \citep{Seifert2013Large-eddyClouds}. The application of these idealized LESs, launched from homogeneous conditions, seems limited to clouds up to around 100 km because spontaneous cloud growth does not significantly exceed 100 km in such setups \citep{Lamaakel2023ComputationalConvection}. 
However, in nature, cloud structures much larger than 100 km are observed. An example of such a cloud structure is a Fish cloud (Figure \ref{fig:Fish-sat}a), resembling a Fish skeleton of 200 to 2000 km \citep{Stevens2020SugarWinds}.  Fish clouds are expected to be large-scale remnants of extratropical disturbances \citep{Aemisegger2021HowRegion, Schulz2021CharacterizationTrades, Villiger2022LagrangianEUREC4A}. Moreover, they are expected to reside upon moisture perturbations \citep{Villiger2022LagrangianEUREC4A}, as confirmed by the Total Water Path (TWP) anomaly in reanalysis data (Figure \ref{fig:Fish-sat}a). 

Given the rationale to understand shallow cumulus patterns (i.e. climate sensitivity), the natural question to ask is: how does such a mesoscale moisture perturbation affect the radiative properties of a cloud field in an idealized LES? To answer this question we conduct four idealized LESs: one launched from homogeneous initial conditions and three containing a sizable mesoscale moisture perturbation. All other simulation aspects, including the domain-mean moisture, are kept constant, and therefore, any difference between the simulations can be attributed to the initial moisture variance. 

\section{The set-up of the simulations with Fish-like large-scale conditions}
\label{sec:setup_LES}
We apply the moisture perturbation to a large-scale environment that favors a large cloud structure, i.e., a Fish cloud. The LES environment contains domain-mean vertical profiles and the sea surface temperature, which are constructed from reanalysis data in regions surrounding Fish structures. These structures are identified using a dataset from a previously applied neural network. The neural network was first based on visible images \citep{Rasp2020CombiningConvection}, then adapted to categorize cloud fields based on brightness temperatures \citep{Schulz2021CharacterizationTrades}, and finally applied to the years 2003-2020 \citep{Eastman2024AMethodologies}. The resulting dataset is used here to find Fish clouds over the tropical Atlantic Ocean east of Barbados (48-58$^{\circ}$ W, 10-20$^{\circ}$ N) during the boreal winter (December, January, February). After detecting the Fish patterns, we extract averages of atmospheric vertical profiles (i.e., $\theta_l$, $q_t$, $u$, $v$, $\omega$) over the 500x500 km$^2$ region around the Fish's  center. This center is defined by the average latitude and longitude of the Fish's coverage. Note, if two Fishes are detected in the domain, the center may lie outside of both structures. Thus, the constructed environment does not strictly represent the environment of any of these two Fishes. However, the simultaneous detection of two Fish patterns suggests a large-scale environment favorable for the presence of Fish structures and therefore this is not considered a limitation.

The resulting large-scale Fish environment serves as initial conditions and large-scale forcings for four simulations conducted with the Dutch Atmospheric Large-Eddy Simulation (DALES) \citep{Heus2010Heus2010Applications}. The environmental profiles are given in the Supplementary material. 
The horizontal dimensions of our doubly periodic simulations span $512 \times 512$ km$^2$ with a grid size of 100 m. There are 175 vertical layers and a vertical spacing of 15 m in the first model level, stretched by 1\% in each level above until a height of 7 km. There is large-scale horizontal advection of heat and moisture based on the velocities and the heat and moisture gradients obtained during the construction of the environment as described above, i.e., $-\overline{u} \frac{\delta \overline{\theta_l}}{\delta x}-\overline{v} \frac{\delta \overline{\theta_l}}{\delta y}$ and $-\overline{u} \frac{\delta \overline{q_t}}{\delta x}-\overline{v} \frac{\delta \overline{q_t}}{\delta y}$, where the bar ($\overline{.}$) represents the domain-average. Moreover, there is vertical advection of heat and moisture with the imposed, domain-averaged subsidence velocity. The RRTMG model is used to calculate the radiative heating rates \citep{Iacono2008RadiativeModels} including the representation of the diurnal cycle. The sources and sinks of precipitation are modeled with a warm microphysics scheme based on \citet{Seifert2006ADescription}, and the cloud-droplet number is fixed at $7 \cdot 10^7$ /m$^3$. 

The Control simulation departs from a horizontally homogeneous state (i.e., the prescribed initial profile). The Perturbed simulations depart from a horizontally heterogeneous state; the initial total-water specific humidity contains a Gaussian perturbation with a maximum amplitude of 1 g/kg. The perturbation is applied to each height below 1100 m. After its application, the specific humidity is reduced in the whole domain such that the horizontally averaged specific humidity remains equal to the Control case. The first Perturbed simulation departs from a 250 km structure. Most shown results arise from this simulation, and therefore this simulation is meant when 'the Perturbed simulation' is described. The other two Perturbed simulations similarly start with an initial perturbation, but we additionally apply the same perturbation at random points throughout their runtime. The aim of this is to idealistically simulate the perturbations that an initial structure, brought into a trade cumulus regime, might encounter within its lifetime. This enables us to compare the effect of an initial mesoscale forcing to the perturbations encountered once a structure has developed. The first of these additional simulations receives pulses that are 250 km in diameter, while the second starts from and receives perturbations of 144 km in diameter. During comparisons, the simulations are referred to as '1 Pulse', 'Random', and 'Random small'. The simulations are initialized at midnight UTC and ran for 72 hours, except for the 'Random small' simulation, which terminated prematurely at 65 hours.

ChatGPT was used in debugging, small plotting improvements, interaction with Snellius, and a brainstorm on moisture variance advection into the trades.

\section{The Perturbed simulations reflects substantially more shortwave radiation than the Control simulation}
\label{sec:develop}
\begin{figure}[h!]
    \centering
    \includegraphics[width=\textwidth]{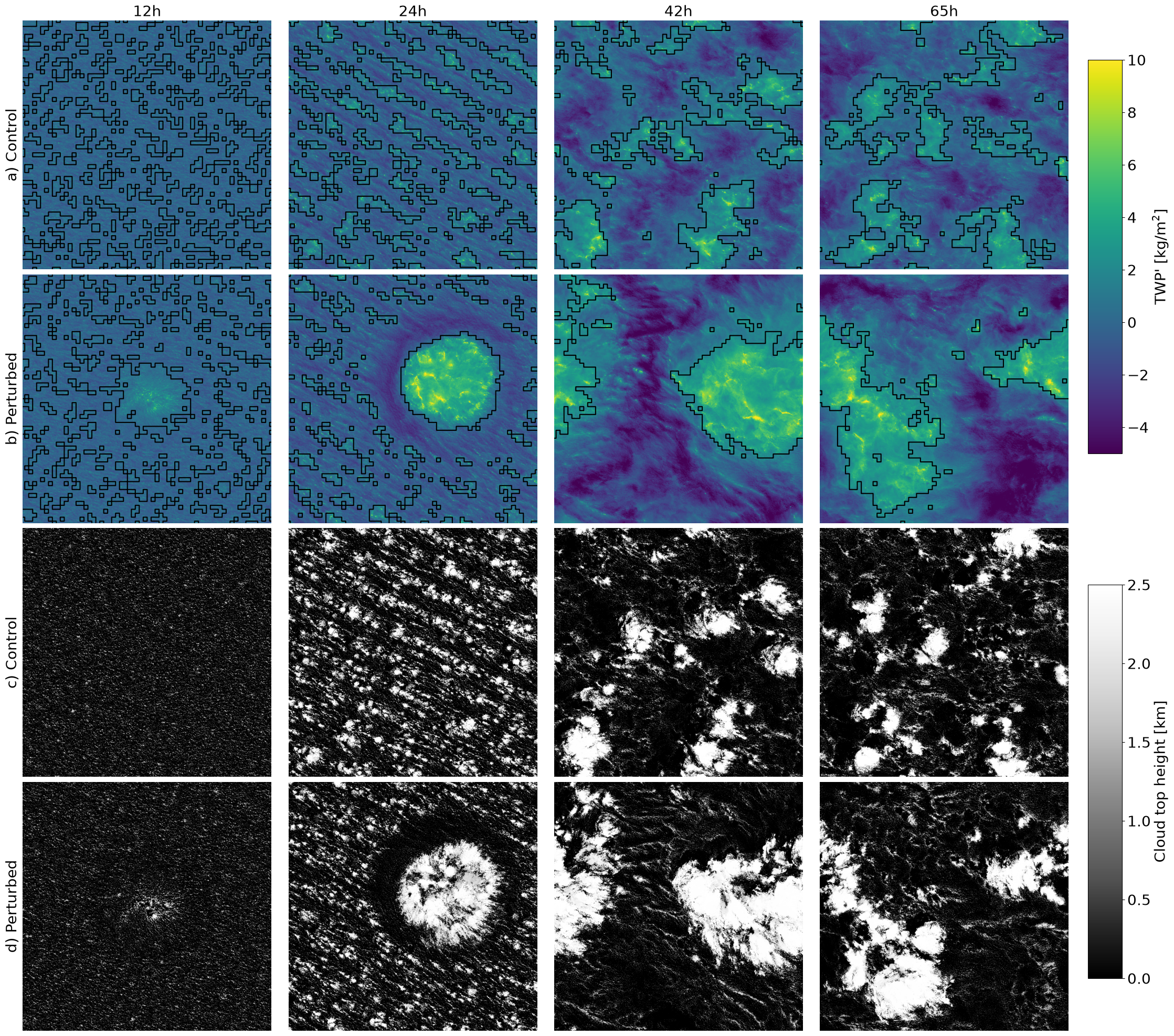}
    \caption{Snapshots of TWP of the Control (a) and Perturbed (b) simulation and cloud top height of the Control (c) and Perturbed (d) simulation at 12, 24, 42 and 65 hours after the initialization. The contour lines in (a) and (b) surround the 25\% wettest mesoscale (8x8 km$^2$) areas.}
    \label{fig:snap_all}
\end{figure}

 In both simulations, trade cumuli initially develop in regions with high TWP (Figure \ref{fig:snap_all}), and these clouds and moist regions subsequently grow in size (i.e. aggregate).
In the Control simulation, this aggregation leads to several 100 km cloud clusters that resemble Flower clouds
(Figure \ref{fig:snap_all}c). This result is in line with \citet{Lamaakel2023ComputationalConvection}, who  find that enlarging the domain from 160x160km$^2$ to 320x320km$^2$ also did not produce structures significantly larger than 100 km.

In the Perturbed simulation, the convection initially develops in the middle of the perturbation and is surrounded by a dry, cloud free band (Figure \ref{fig:snap_all}b and d). Crucially, the moisture is not rapidly redistributed horizontally, and the simulation does not return to a  moisture pattern that matches the Control case within 72 hours (Figure \ref{fig:snap_all}a and b). Instead, the initial moisture perturbation and its shallow convection grow in horizontal size and amplitude (Figure \ref{fig:snap_all}b and d). Eventually, the cloud attains an elongated shape and thus begins resembling a Fish cloud (Figure \ref{fig:Fish-sat}a, and Figure \ref{fig:snap_all}d 42h). Moreover, the cloud reaches a length of around 300 km, which is larger than the diameter of the imposed moisture perturbation (250 km). The other Perturbed simulations give qualitatively the same results. This emphasizes that mesoscale feedbacks can enhance large mesoscale moisture perturbations, and consequently generate visually very different cloud structures over the time scales that clouds traverse the trades than cloud structures in situations without such perturbations.

\begin{figure}[h!]
    \centering
    \includegraphics[width=\linewidth]{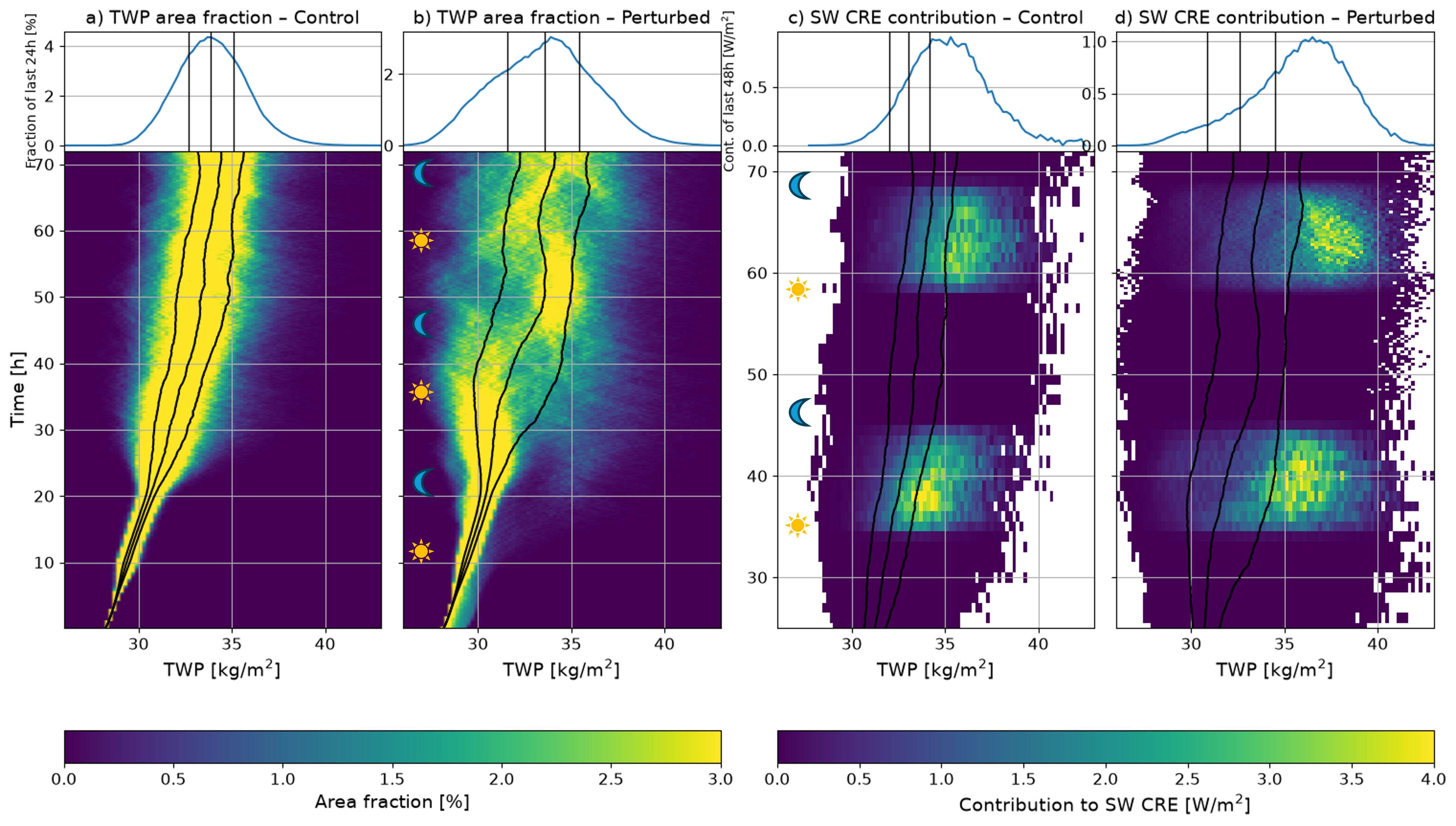}
    \caption{a) and b) the area fraction of bins of mesoscale (8x8 km$^2$) TWP. The colour at a given point (e.g. TWP = 30 kg/m$^2$, time =  20 h) provides the area fraction [\%] covered by 8x8km$^2$ cells having a TWP of 30 kg/m$^2$ at 20 hours. The sun indicates the sunrise and the moon the sunset. Atop the area fraction averaged over the last 24 h. c) and d) the contributions to the SW CRE [W/m$^2$], i.e., the SW CRE of a bin weighted by the area fraction of the bin. Atop the average over the last 48 h.  a) and c) are the Control case and b) and d) the Perturbed case. The black lines resemble the 25$^{th}$, 50$^{th}$ and 75$^{th}$ TWP percentile.}
    \label{fig:Hov}
\end{figure}

The evolution of the TWP field is quantified by first subsampling the field to a resolution of 8x8 km$^2$, since the TWP variance is dominated by scales larger than 8 km. Thereafter, the TWP is binned and each bin's area fraction [\%] is calculated (colored panels in Figures \ref{fig:Hov}a and b). 

The resultant TWP distributions develop very differently between the Control and the Perturbed simulation. The TWP distribution of the Control case shifts towards more moisture during the full simulation and widens around 20 hours. In the Perturbed case, the TWP distribution also widens around 20 hours and develops a clear diurnal cycle thereafter. The first day is interpreted as the spin-up of the simulations and therefore in the remainder of this report, values are averaged over the last two simulation days, where not stated otherwise. The median TWP does not differ much between the simulations; averaged over the last 24 hours, it is 33.9 kg/m$^2$ in the Control and 33.6 kg/m$^2$ in the Perturbed simulation (middle black line in Figure \ref{fig:Hov}a and b). However, the interquartile range is smaller in the Control (32.7-35.1 kg/m$^2$) compared to the Perturbed case (31.6-35.4 kg/m$^2$) (left and right black lines in Figure \ref{fig:Hov}a and b). This means that the moistest quartile is dryer in the Control compared to the Perturbed case. To give a physical interpretation to this moistest quartile, we have drawn its boundaries with black contour lines in Figure \ref{fig:snap_all}c and d. The encircled columns (henceforth $Q_4$) encapsulate the Flower clouds in the Control simulation (Figure \ref{fig:snap_all}c) and the Fish cloud in the Perturbed simulation (Figure \ref{fig:snap_all}d). The other Perturbed simulations also have an increased interquartile range compared to the Control case and a similar TWP median. The random pulses case develops a bimodal TWP distribution.

The net radiative effect is 6.8 W/m$^2$ larger in the Perturbed compared to the Control simulation. This is much larger than the net radiative effect of other mesoscale processes. The radiative effect of cold pools and of symmetrical growth are quantified by denial experiments in idealized LES ensembles. Notably, our value lies outside their ranges, i.e. approximately -0.5 to 4.7 W/m$^2$ for cold pools (average: 1.88 W/m$^2$) \citep{Alinaghi2025WarmingBudget} and -4 to 2 W/m$^2$ for symmetrical growth (average: -0.5 W/m$^2$) \citep{Janssens2025SymmetrySimulations, McCoy2025DoesCluster}. Our radiative cooling can also be compared to the radiative response of shallow clouds to the large-scale variability, which can be determined by the current Cloud Controlling Factor (CCF) variability and its influence on the radiation. Our cooling is larger than  such large-scale influences in observations (Figure 8 in \citet{Scott2020ObservedOceans}).

The enhanced radiative cooling (6.8 W/m$^2$) is caused by a stronger shortwave (SW) cloud radiative effect (CRE) (6.6 W/m$^2$), by a stronger LW clear sky (cs) radiation (0.6 W/m$^2$), and by a weaker LW CRE (-0.4 W/m$^2$). This result is robust to a different simulation set up, where the air mass encounters several moisture perturbations in its lifetime. Such random perturbations increase the SW CRE and LW cs response, but reduce the LW CRE difference (Figure \ref{fig:cf_ql_time_prof}b). However, the SW and LW CRE are not doubled and thus the initial perturbation has more impact on the CREs than the sum of all later-applied perturbations. This does not hold for the LW cs; the impact of the later-applied perturbations on the LW cs is larger than the impact of the initial perturbation. Smaller random pulses decrease all radiation differences (Figure \ref{fig:cf_ql_time_prof}b). The difference between 1 single pulse and random pulses is smaller than the difference between random large pulses and random small pulses. Therefore, in our idealistic setup, the horizontal size seems to be more important than the frequency of the pulses. We refrain from providing a quantitative relation between the radiation response and the size and frequency of the externally applied moisture variance, since we argue this can best be done with observational/reanalysis data or an ensemble of simulations, for which we do not have the computational time.

The largest contribution to the net radiative cooling, the SW CRE, is significantly larger throughout the simulation, with a peak difference of 35 W/m$^2$ at 40 hours (Figure \ref{fig:cf_ql_time_prof}a). Figure \ref{fig:Hov}c and d link the local SW CRE back to the TWP distribution, by showing the contribution of each TWP bin. This contribution is calculated by multiplying the area fraction of a TWP bin by the average SW CRE of this bin. Most of the SW CRE signal clearly originates from locations with large TWP, i.e., $Q_4$: the Flower clouds and Fish cloud respectively in the Control and Perturbed simulation. This is even more clear in the contribution to the SW CRE averaged over the last 48 simulated hours (curves atop the coloured panels). These curves quantify that 67\% of the SW CRE is caused by $Q_4$ in the Control case and 72\% in the Perturbed case. Thus, the wettest bins cause most of the SW CRE. This means that the longer moist tail that develops in the Perturbed simulation is responsible for generating more reflective clouds.

\section{The Perturbed simulation reflects more sunlight because its cloud cover is larger}
\label{sec:CC}
 
\begin{figure}[h!]
    \centering
    \includegraphics[width=0.9\linewidth]{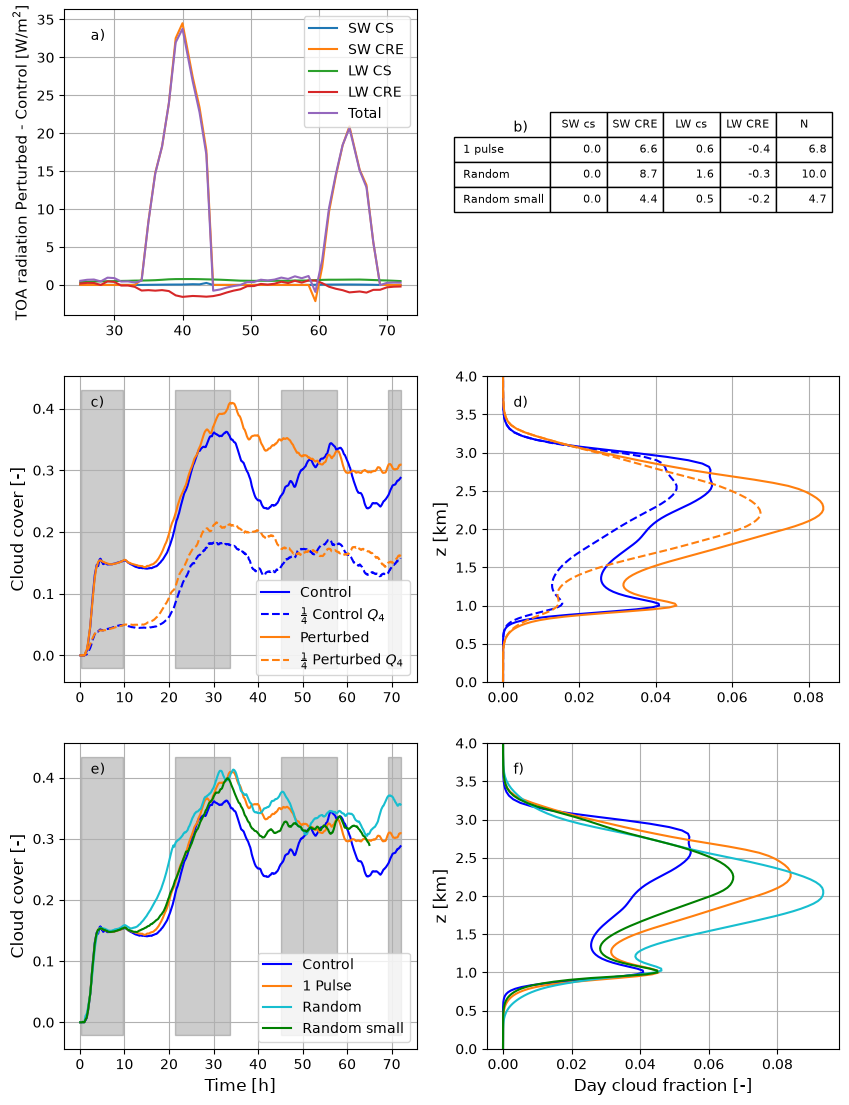}
    \caption{a) The difference in top of the atmosphere (TOA) radiation between the Perturbed and Control case. b) The difference in TOA radiation compared to the Control simulation, for all three Perturbed simulations. c) and e) the time-evolution of the cc, and d) and f) profiles of the cloud fraction, daytime averaged over the last 48h. c) and d) containing the contributions of the 25\% wettest columns (i.e. the $Q_4$ area-weighted averages) and e) and f) containing all Perturbed simulations. The gray rectangles in c) and e) represent night times.}
    \label{fig:cf_ql_time_prof}
\end{figure}

The Perturbed simulation has a larger SW CRE compared to the Control simulation not because of cloud brightening, but because its daytime cloud cover (cc) is larger (33\%) compared to Control (28\%).The enhanced cc can be studied in more detail by visualizing its time evolution and the vertical profile of the cloud fraction. The cc is similar in both simulations until 30 h (Figure \ref{fig:cf_ql_time_prof}c). After 30 h the cc
is larger in the Perturbed simulation compared to the Control simulation, with the difference most pronounced during daytime. This large difference during daytime is to a large extent caused by the cc dropping in the Control simulation upon sunrise (34 h and 58 h), while the cc in the Perturbed simulation is much less sensitive to the daily cycle of radiation. In the random cases, the larger SW CRE is also caused by a larger cc (Figure \ref{fig:cf_ql_time_prof}e), where an increased perturbation size and frequency enlarge the cc.

The difference in cc is mainly caused by clouds between 1.5 and 2.5 km altitude (Figure \ref{fig:cf_ql_time_prof}d), which seem to be stratiform clouds (Figure \ref{fig:snap_all}b and d) flowing from the tops of the mesoscale convective systems. This also holds for the random cases (Figure \ref{fig:cf_ql_time_prof}f). Having more stratiform clouds in all Perturbed simulations implies that the amount of these higher clouds increases due to the initial mesoscale moisture perturbation. This is in line with \citet{Dauhut2026SensitivityHeterogeneities}, who also find that moisture heterogeneity leads to more of these higher elevated clouds, in a Flower case study.

These higher clouds mostly occur in $Q_4$ (dotted lines in Figure \ref{fig:cf_ql_time_prof}d), i.e., in the Fish cloud and Flower clouds of the Perturbed and Control simulation respectively. Both the Fish cloud and the Flower clouds contain more high clouds than cloud base clouds, in line with observations \citep{Vial2021OnTrades}. Since the high clouds are concentrated in $Q_4$, their cloud fraction is more variable than the cloud base cloud fraction, in line with previous LESs \citep{Bretherton2017UnderstandingSimulation} and observations \citep{Nuijens2014TheTrades}.

\section{Summary, discussion and outlook - potential implication for the shallow cloud feedback}
\label{sec:disc}
We compare three 'Perturbed' simulations with a single or several moisture perturbations to a 'Control' simulation in idealized $512 \times 512$ km$^2$ LESs. The perturbations grow into a self-sustaining system with a large cloud. The imposed mesoscale moisture variance leads to a different vertical and horizontal structure of the cloud and moisture. The main difference, a larger stratiform cloud cover, leads to a much larger SW CRE compared to the Control simulation. Accordingly, the Perturbed simulations absorb 5-10.0 W/m$^2$ less net radiation compared to the Control simulation. This difference is larger than differences due to other mesoscale processes \citep{Alinaghi2025WarmingBudget, Janssens2025SymmetrySimulations} and due to current changes in the large-scale environment \citep{Scott2020ObservedOceans}.

The enhanced radiative cooling may influence the climate sensitivity associated with patterning of trade-wind cumulus clouds. If the advection of moisture variance into the trades becomes less (more) frequent with warming, the frequency of Fish-like clouds could decrease (increase), which would increase (decrease) the shallow cloud feedback, currently estimated to be between -0.5 and 0.5 W/m$^2$/K \citep{Myers2021ObservationalSensitivity}. If our idealized LES-assessed moisture variance effect on the radiation is in the right ballpark, a change in the frequency of Fish-like clouds of merely 7.4\% (6.8 $\cdot$ 7.4\% = 0.5) would bring the feedback outside its currently estimated range. 

Given the idealized setup of our simulations, these numbers (6.8 W/m$^2$ and 7.4\%) are not intended as a quantification of the impact of mesoscale moisture variance on the radiation field of trade cumuli. However, the numbers are large enough to indicate that moisture variance advected into the trades could constitute a potentially important, overlooked contribution to the cumulus feedback. Here, we will discuss this potential impact by 1) hypothesizing how advection of mesoscale moisture variance into the trades could become more or less frequent in a warmer world, 2) hypothesizing which quantity of the moisture variance matters for the final cloud size and 3) hypothesizing how future work could account for moisture variance in a current method to estimate cloud feedback. Finally, we will provide a conclusion.

The frequency of moisture variance brought to the trades could change with warming. \citet{Aemisegger2021HowRegion, Villiger2022LagrangianEUREC4A} connect Fish structures to Rossby Wave Breaking (RWB). Yet the response of RWB to global warming is uncertain. According to Global Climate Models, their frequency will decrease \citep{Tahvonen2025ResponseCover, Barnes2012DetectionChange}. However, in reanalysis data, RWB frequency has increased in a 35-year trend when identified at isentropic surfaces above 320 K \citep{Jing2018RossbyHemisphere}. Therefore, both a decrease as well as an increase in the occurrence of advected moisture variance from the extratropics may be plausible. The first could lead to a negative contribution of Fish clouds to the climate feedback and the second to a positive contribution. Thus, for quantifying the cumulus cloud feedback induced by imposed mesoscale moisture variance, consensus should be reached on the response of RWB to climate change. Moreover, other possible origins of moisture variance (e.g. remnants of deep convection?) should be mapped. 

In the reasoning above, we assumed the occurrence of advected moisture variance to correlate to the occurrence of Fish clouds. However, this correlation is not quantified yet and therefore we discuss here a quantity of the moisture variance that seems to be relevant in this correlation, i.e., the horizontal scale. In our simulation set-up, a large (300 km) Fish-like cloud is formed if the simulation is initiated with mesoscale moisture variance, while smaller (100 km) Flower-like clouds spontaneously arise. However, in another case-study, a Flower cloud is formed in the presence of moisture variance, while an unorganized cloud field spontaneously arises \citep{Dauhut2026SensitivityHeterogeneities}. While the horizontal size of our moisture perturbation is 250 or 144 km in diameter, their horizontal size is approximately 60 km (their Figure 2a and b). Thus, our sizes are above the final size of spontaneous cloud growth (100 km) and theirs is below this size. Therefore, we wonder if there exists a critical threshold of moisture perturbations between clouds growing to 100 km and clouds growing larger. A similar horizontal scale helps to predict the occurrence of self-aggregation in deep convection in cloud-resolving model simulations \citep{Silvestri2026PredictingLength-scales}. The correlation between moisture variance size and the final cloud size, as well as the correlation between the final cloud size and the radiation remains to be quantified.

Such quantifications might help to improve current estimates of the trade-wind cumulus cloud feedback. This feedback is currently estimated using CCFs in a two-step manner: 1) the response of low-cloud induced radiative fluxes to changes in CCFs are modeled based on observations and reanalysis data and 2) the changes in these CCFs with warming are obtained from Global Climate Models \citep{Myers2021ObservationalSensitivity}. 
The first step is a multilinear regression model on a spatial scale similar to ours and explains 37\% of the variance \citep{Scott2020ObservedOceans, Myers2021ObservationalSensitivity}. Based on our study and the results of \citet{Dauhut2026SensitivityHeterogeneities}, we expect this model to improve when adding (the horizontal size of) mesoscale moisture variance to the CCFs.

Overall, adding a mesoscale moisture perturbation to a homogeneous trade-wind environment seems to influence physical processes of shallow convection, which modify the SW CRE through changes in stratiform cloud amount. With global warming, the advection of moisture variance into the trades could change. Our results indicate that such changes would likely give significantly changed CRE of cumulus clouds, and in turn affect the trade cumulus cloud feedback. Therefore, this work highlights the importance of understanding how mesoscale heterogeneities imposed on the trade-wind regions will change with warming, and how they will affect shallow convection, its organization and its radiative effects. In fact, it opens a different view on the cloud feedback, which is currently a linear sum of feedbacks from individual cloud types, each considered separately \citep{Sherwood2020AnEvidence}. Our work emphasizes that connecting regimes of cloud feedbacks is an open but important research field.

%
%

\section*{Open Research Section}
The version of DALES, the postprocessing code, the data and the code to generate the plots are publicly available in \citet{vanDriel2026SoftwareSoftware}.

\section*{Conflict of Interest declaration}
“The authors declare there are no conflicts of interest for this manuscript.”

\section*{Acknowledgements}
The simulations are carried out on the Dutch national e-infrastructure with the support of the SURF Cooperative and funding from the Nederlandse Organisatie voor Wetenschappelijk Onderzoek (grant no. NWO-2025.010). MJ acknowledges the project Connecting the Clouds with file number VI.Veni.242.017 of the research programme Veni which is financed by the Dutch Research Council (NWO) under the grant https://doi.org/10.61686/TJSFP13491.

\section*{Supplement}

\begin{figure}[h!]
    \centering
    \includegraphics[width=\linewidth]{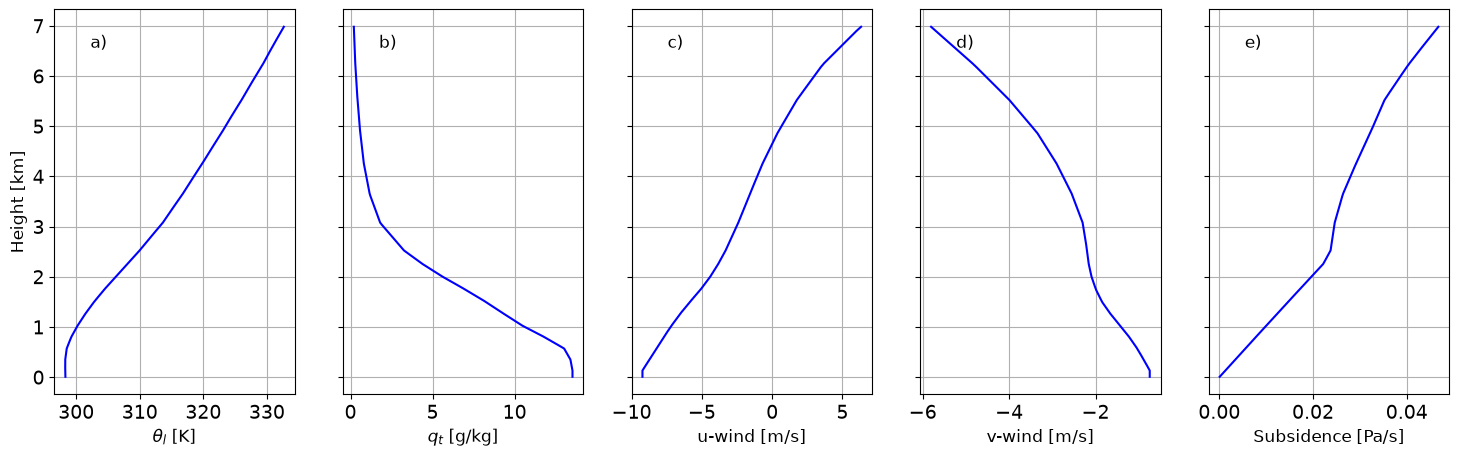}
    \caption{Initial profiles of the simulations.}
    \label{fig:init}
\end{figure}

\bibliographystyle{apacite}
\bibliography{references}  

\end{document}